\documentclass{article}

\PassOptionsToPackage{square, numbers}{natbib}
\usepackage[final]{neurips_mic_2021}

\usepackage[utf8]{inputenc} % allow utf-8 input
\usepackage[T1]{fontenc}    % use 8-bit T1 fonts
\usepackage{hyperref}       % hyperlinks
\usepackage{url}            % simple URL typesetting
\usepackage{booktabs}       % professional-quality tables
\usepackage{amsfonts}       % blackboard math symbols
\usepackage{nicefrac}       % compact symbols for 1/2, etc.
\usepackage{microtype}      % microtypography
\usepackage{xcolor}         % colors

\usepackage{graphicx}
\usepackage{verbatim} %comments
\usepackage{siunitx} %scientific notation (for p values)
\graphicspath{ {./figures/} }
\usepackage{subcaption}

\title{The role of joint utility and pragmatic reasoning in cooperative communication}
\author{%
Yiling Yun\\
  Department of Statistics,
  University of California, Los Angeles, USA 90024\\
  \texttt{yiling.yun@g.ucla.edu} \\
  \And
  Stephanie Stacy\\
  Department of Statistics,
  University of California, Los Angeles, USA 90024\\
  \texttt{stephaniestacy@g.ucla.edu} \\
  \AND
  Tao Gao\\
  Department of Statistics and Communication,
  University of California, Los Angeles, USA 90024\\
  \texttt{tao.gao@stat.ucla.edu} \\
}

\begin{document}

\maketitle

\begin{abstract}
  Humans are able to communicate in sophisticated ways with only sparse signals, especially when cooperating. Two parallel theoretical perspectives on cooperative communication emphasize pragmatic reasoning and joint utility mechanisms to help solve ambiguity. For the current study, we collected behavioral data which tested how humans select ambiguous signals in a cooperative grid world task. The results provide support for a joint utility reasoning mechanism. We then compared human strategies to predictions from Rational Speech Acts (RSA), an established model of language pragmatics. 
\end{abstract}

\section{Introduction}
\label{intro}
Humans are highly cooperative. We see evidence in early humans, infants, and, on a larger scale, charitable organizations \citep{tomasello2014natural, warneken2006altruistic, sugden1993thinking}. This cooperative foundation allows humans to leverage communication to facilitate increasingly sophisticated coordination. In fact, one philosophical school of thought treats language as a social tool for cooperation \citep{vygotsky2012thought, clark1996using}. Even for complex tasks, we often only need sparse and ambiguous signals to successfully communicate. For example, imagine one person is cooking and asks his friend to get the pot on the stove top. Seeing his friend about to reach the pot with bare hands, the person says “hot.” His friend might quickly withdraw his hands and get some gloves for the pot. Here the speaker expects his friend to understand the signal without specifying if “hot” describes the pot or the weather, and why it means “get some gloves” instead of only describing the state of the pot. These types of ambiguous signals are often understood because the receiver is an intelligent “inference-making machine” \citep{sacks1985inference} who expects their partner to also be intelligently reasoning about the situation. This example highlights that, by assuming our partner is cooperative, a signal’s meaning is restricted to things that are helpful and relevant to the current situation. The current study aims to better understand this phenomenon by testing how humans use sparse language to communicate in a physical cooperative task and examining to what extent human behavioral evidence aligns with current models of cooperative communication.

Two independent but complementary lines of reasoning consider cooperation in communication. The first is pragmatic reasoning: interpreting the meaning of a signal by considering its context. Grice helped develop this field by placing language in the context of cooperation, which allows both interlocutors to assume their partner is communicating in the most straightforward manner possible. That is, the signals we send should be truthful towards the goal, relevant to the current situation, and as brief as possible \citep{grice1975logic}. When both communicators act under this assumption, a signal with more than one literal meaning is often resolved by pragmatic reasoning. Computationally, Rational Speech Acts (RSA) is a well-established and successful model of pragmatic reasoning in language \citep{frank2012predicting, goodman2016pragmatic, kao2014formalizing, yoon2017won}. The original model simulated a pragmatic listener, and a natural extension of this work was to model communication from the pragmatic speaker's perspective \citep{monroe2015learning}. 

The second perspective on cooperative communication is grounded in interactive tasks: communication here is utility-driven, from a shared agency perspective. Cooperators share a joint goal that they both recognize and work towards -- critically -- as a “collective body” \citep{gilbert2013joint} or a team. If there is a rule of action that is good for the team, each member will follow this rule \citep{sugden1993thinking}. In order to coordinate, cooperators may take a “bird’s-eye view” \citep{tomasello2010origins} or a view from “nowhere in particular” \citep{nagel1989view}. Even children coordinate using shared agency which has been demonstrated with empirical evidence of flexible role-reversal and awareness of joint commitment in cooperation \citep{fletcher2012differences, warneken2006cooperative}. 

Joint utility comes into play when we combine the shared agency perspective with utility calculus. It is well established that people act rationally to maximize individual utility \citep{jara2016naive, liu2017ten, baker2009action}. In cooperative settings, people are able to consider how to maximize joint utilities, which has supporting empirical evidence. When cooperating with a partner towards a goal, people considered the utility of joint actions over either agent’s individual utility \citep{torok2019rationality}. This cooperative logic can also be found in infants. When an adult made an ambiguous request to infants in a visual context, they could resolve ambiguity by interpreting the request in a way that maximizes joint utility \citep{grosse201021}. In addition to empirical evidence, computational approaches have successfully modeled how to coordinate once the high-level decision of cooperation is made \citep{kleiman2016coordinate}.

Based on the two theoretical perspectives on how people communicate in cooperative settings, we are interested in how these components manifest themselves in a cooperative task. We took the RSA reference game and grounded it in a visual context with utility considerations, similar to Grosse’s experiment \citep{grosse201021}, but with more alternatives that made the task more complicated. Our study also involved a more complex utility structure that allowed both cooperators to act and changed their responsibilities through a barrier in the visual context. We collected human data and compared it with simulations from RSA to see to what extent pragmatics reasoning and joint utility match human signaling behavior.

\section{Methods}
\label{methods}
\begin{figure}
     \centering
     \begin{subfigure}[b]{0.4\textwidth}
         \centering
         \includegraphics[width=\textwidth]{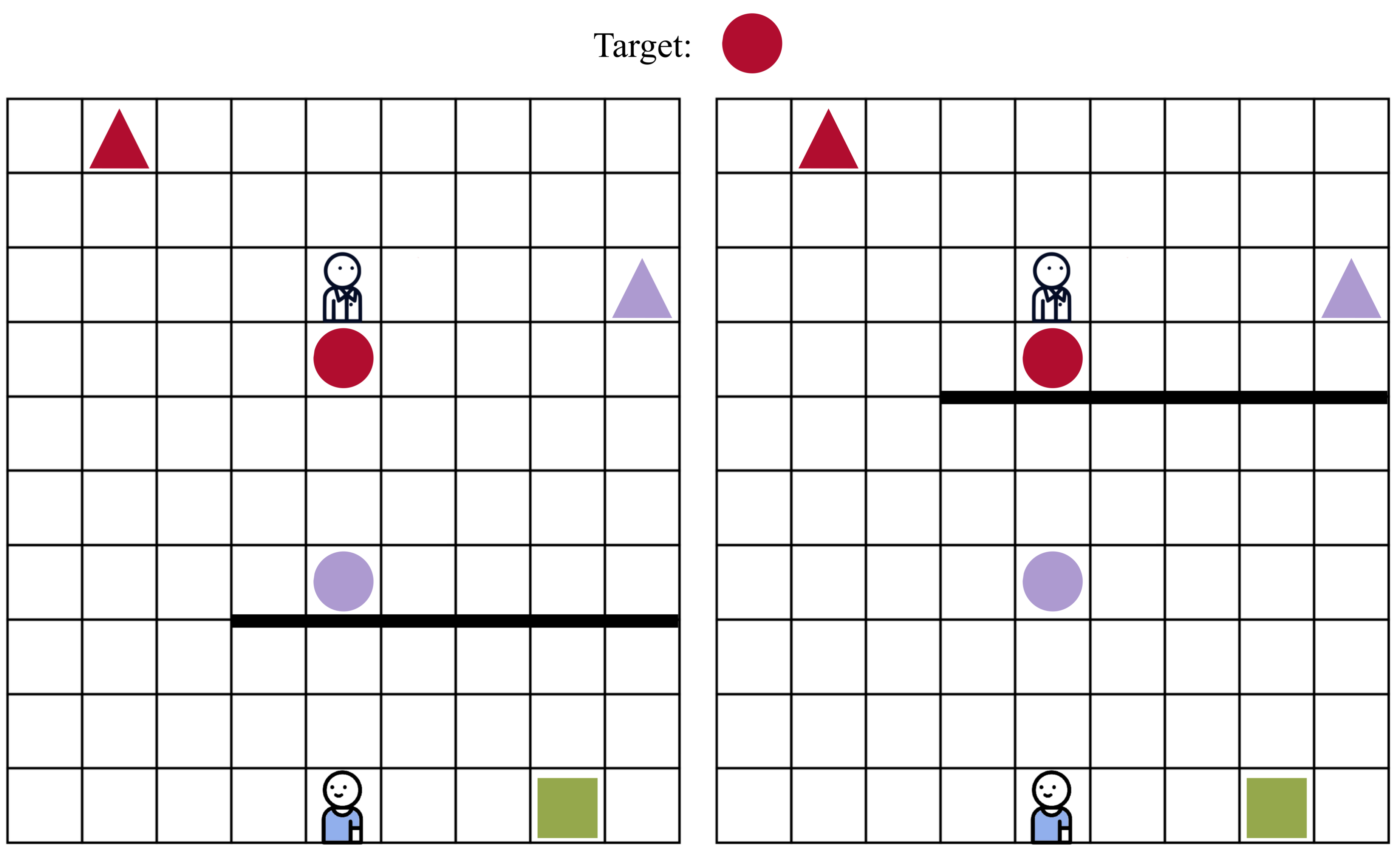}
         \caption{$Simple$}
     \end{subfigure}
     \hfill
     \begin{subfigure}[b]{0.4\textwidth}
         \centering
         \includegraphics[width=\textwidth]{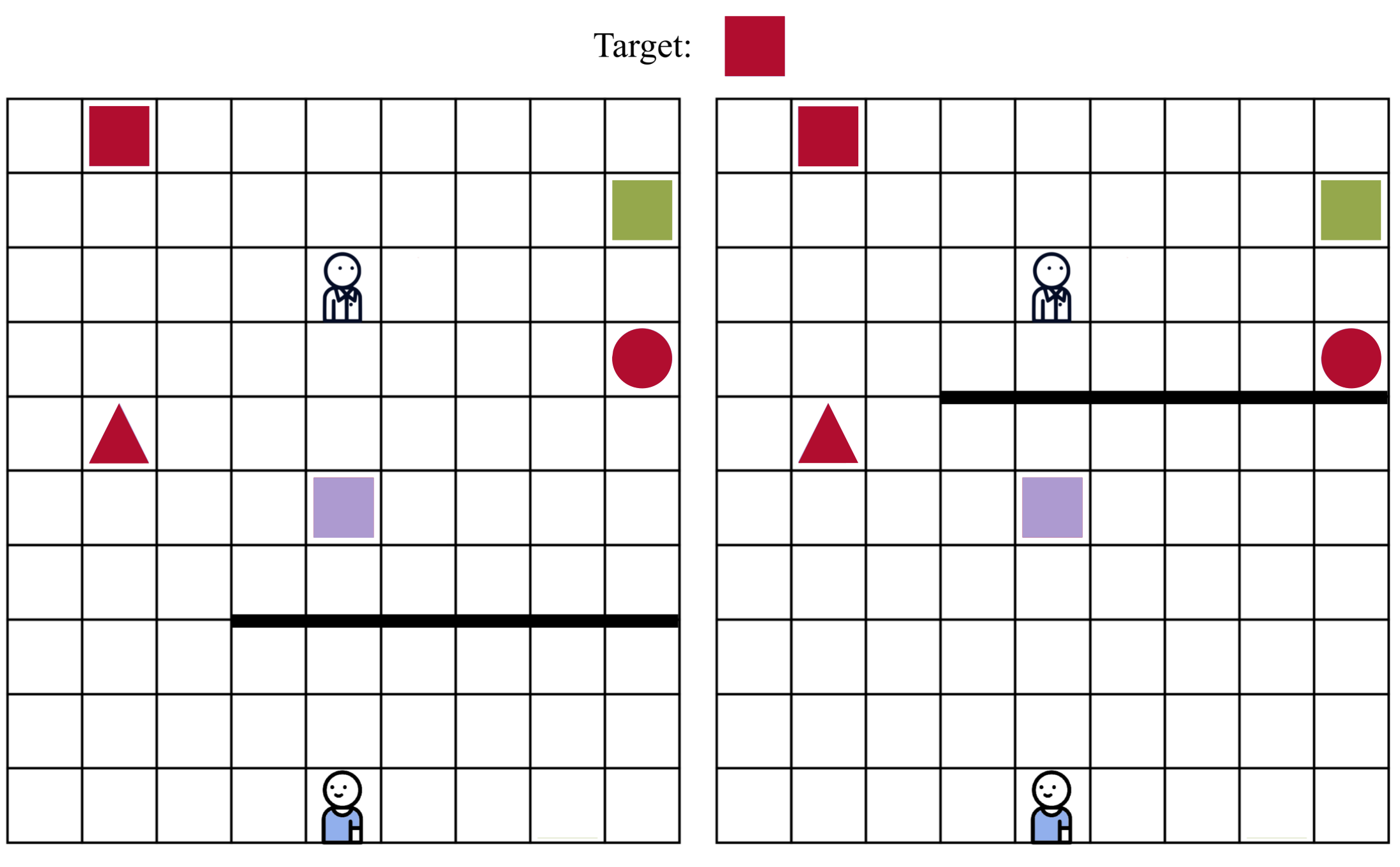}
         \caption{$Difficult$}
     \end{subfigure}
     \caption{\textbf{Paired example trials.}}
  \label{fig:gridworldexample}
\end{figure}

Participants played a cooperative communication game in a grid world environment where their goal was to reach a target in the fewest steps (Figure \ref{fig:gridworldexample}). All trials have five items with two features of color (red, purple, or green) and shape (triangle, circle, or square). Participants were always the signaler in blue, who had a full view of the grid world and knew the target. The receiver in white was pre-programmed to respond, but participants were told that the receiver was intelligent and motivated to help. In each trial, the task was for participants to decide whether to go to the target themselves or send a signal describing a single feature of an item. If either agent reached the correct target, the participant would receive a reward of \$0.40. Otherwise, the reward was \$0.00. The cost of each step was -\$0.05. There was no cost for sending a signal, but only one signal was allowed in each trial. The target was always near the receiver in experimental conditions so that the optimal utility could only be achieved with communication.

% Based on the utilities for either actor to reach each grid when the barrier changes its location, we partitioned the grid world into four sections (see Figure \ref{fig:partitioned_gridworld}). 

% \begin{figure}[ht]
%   \centering
%   \includegraphics[width=0.30\linewidth]{figures/partitioned_gridWorld.png}
%   \caption{\textbf{Partitioned grid world.} The green grids were always the receiver’s responsibility. The blue grids were always the signaler’s responsibility. The red grids shifted from the receiver’s responsibility to the signaler’s responsibility when the barrier towards the receiver. The white grids were designed to not contain any item.}
%   \label{fig:partitioned_gridworld}
% \end{figure}

The experiment followed a 2 x 2 within-subject design. The first independent variable was the barrier located near either the receiver (R) or the signaler (S). Moving the barrier changed who was responsible for the critical item between the two barriers. The second independent variable was the condition of the trial setup. Using only language, the Simple trials could not be solved using pragmatic reasoning, while the Difficult trials could be solved using pragmatic reasoning. Together, we examined how the shared-agency perspective could be integrated with pragmatic reasoning. Specifically, in the Simple trials, when the barrier was near R, the critical item became more costly for R to reach. Looking at the remaining three items that were near R, there was one feature that referred to the target without ambiguity, which should be the optimal signal. In Figure \ref{fig:gridworldexample}(a), the purple circle was the critical item, and "circle" became the optimal signal when the barrier was near R. In the Difficult trials, either feature of the target unambiguously referred to the target using pragmatic reasoning. When the barrier shifted towards R, the critical item became farther from R. In this case, one feature of the target was less ambiguous because it had less literal alternatives among the items that were still near R. In Figure \ref{fig:gridworldexample}(b), when the barrier was near R, "square" became the optimal signal because it could refer to only two items since the purple circle was farther from R, while "red" could refer to three items. In addition to the experimental conditions, participants saw a control designed to test whether they could rationally maximize utility. Here, one feature of the target uniquely referred to the target; however, when the barrier was moved towards R, participants could earn a higher utility by going themselves instead of signaling. Each combination of barrier locations and conditions contained six grid world environments, resulting in 36 trials which were presented in a random order. We investigated two dependent variables: signaling behavior and reaction time. 

\section{Results}
\label{results}
As RSA is typically intended for language only, we extended it to include an individual utility consideration that occurred after inference. In our simulations, the listener made inferences based only on language, while the pragmatic speaker considered the action utility of how many steps to take to reach each item as well as the potential rewards and then weighted by the listener's inferences. To be specific, given the target, the simulated signaler $S$ tended to take the action $a_S$ that maximized the expected utility:

\begin{equation}
    \label{RSA_speaker}
    P_{S}(a_S|target) \propto e^{(\lambda \mathbb{E}[U(a_S, target)])}
\end{equation}. 

Here, $a_S$ could be going itself or sending out a signal. If the signaler went by itself, the utility evaluated the action costs of walking to the target and the reward. The expected utility was modeled as follows:

\begin{equation}
    \mathbb{E}[U(do, target)] = U(do, target)
\end{equation} 

If the signaler sent out a signal, the utility evaluated the action costs of the receiver $R$ walking to an item and the potential reward, where $a_R$ was the item that the receiver reached. The expected utility was modeled as follows:

\begin{equation}
    \mathbb{E}[U(signal, target)] = \mathbb{E}_{P(a_R|signal)}[U(a_R, target)]
\end{equation}

The receiver distribution $P(a_R|signal)$ came from $P_{R}(goal|signal)$, which was modeled by $P_{R}(goal|signal) \propto P_{litS}(signal|goal)P(goal)$, where $P_{lit S}(signal|goal)$ was a literal signaler boolean function returning 1 when the $signal$ was consistent with the $goal$. The rationality term $\lambda$ in Equation \ref{RSA_speaker} measured how rational the receiver was thought to be. We tried values from 1 to 10 in increments of 1 and used $\lambda = 4$.

\begin{figure}[ht]
    \centering
    \includegraphics[width=0.95\linewidth]{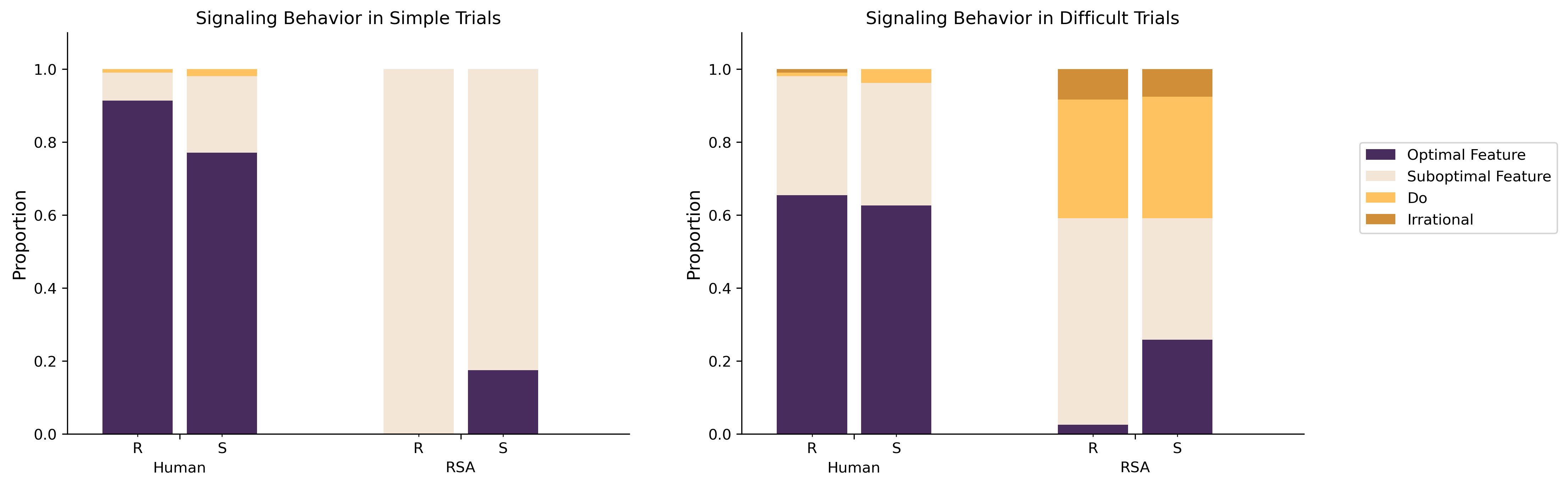}
    \caption{\textbf{Signaling behavior.} "Optimal Feature" was the optimal signal when the barrier was near R. "Suboptimal Feature" was the other feature of the target. "Do" was when the signaler walked by itself. "Irrational" was when the signal was not a feature of the target.}
    \label{fig:comm_human}
\end{figure}

We performed analysis on 650 trials from participants and 720 trials from RSA simulations. In the Simple trials, participants were significantly more likely to communicate the optimal signal when the barrier was near R (91.4\% in R, 77.1\% in S, $p = \num{4.45 e -3}$) (Figure \ref{fig:comm_human}), while RSA speaker was significantly less likely to communicate the optimal signal when the barrier was near R (0.00\% in R, 17.5\% in S, $p = \num{1.60 e -6}$). In the Difficult trials, there was no significant difference between communicating the optimal signal or not in participants (65.4\% in R, 62.6\% in S, $p = 0.669$), while the RSA speaker was significantly less likely to communicate the optimal signal when the barrier was near R (2.50\% in R, 25.8\% in S, $p = \num{2.00 e -7}$).

\begin{figure}[ht]
  \centering
  \includegraphics[width=0.48\linewidth]{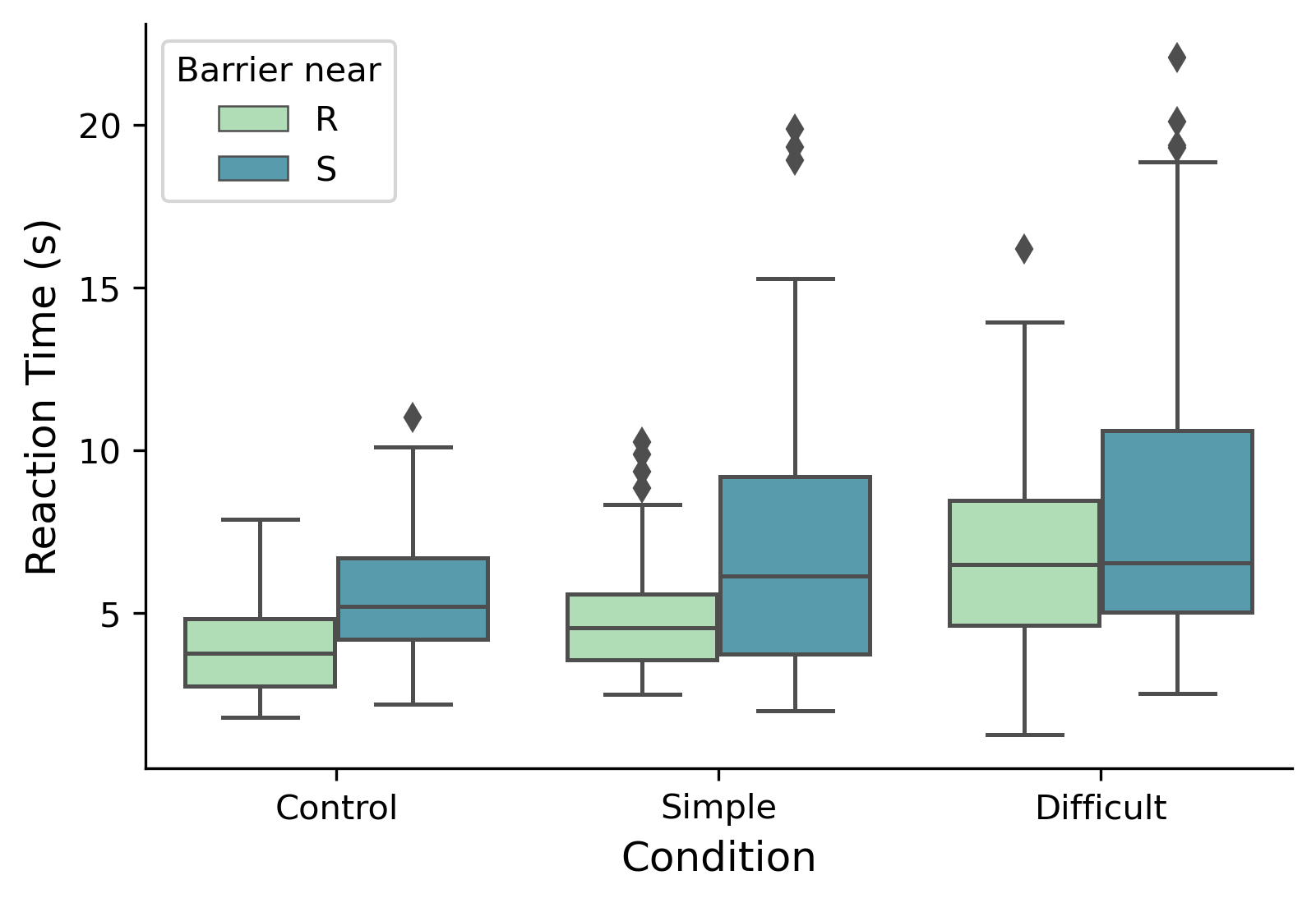}
  \caption{\textbf{Reaction time.} Measured from participants seeing the trial to making a decision.}
  \label{fig:rt_human}
\end{figure}

Furthermore, the reaction time was significantly shorter when the barrier was near R both in the Simple trials ($\mu_R  = 4.85$, $\mu_S = 7.01$, $p =  \num{5.63 e -7}$) and in the Difficult trials ($\mu_R  = 6.94$, $\mu_S =8.15$, $p =  0.0124$) (Figure \ref{fig:rt_human}). Participants' performances were uniform across trials. We did not see evidence of learning or fatigue effect. From the debriefing survey, 95\% of the participants rated the receiver’s performance as "Good" or "Excellent" (1 participant rated "Average"). 57.1\% of the participants reported that they were, to some or greater extent, motivated by the potential money reward.

\section{Discussion}
\label{discussion}
This task provides evidence that humans do consider joint utility as a heuristic for how to disambiguate signals. In the grid world environment, when the barrier is near the receiver, joint utility reasoning allows agents to exclude some items from consideration. Participants were sensitive to this change as evidenced by their significant difference in signaling behavior when the barrier location changed in the Simple trials. Since the receiver's path to the target was not affected by the barrier, the utility for the receiver to reach the target remained unchanged. Therefore, if human participants only considered the individual utility of the receiver, we would not observe this difference. Behavioral data support two-step reasoning in this task: first, joint utility constrains considered items to those that are the receiver's responsibility, and second this constraint allows one signal to identify the target without ambiguity. The supporting evidence is participants' strong tendency to send the optimal signal when the barrier was near the receiver. Furthermore, joint utility reasoning makes the task easier, supported by the significant decrease in reaction time. To understand the misalignment between RSA simulations and human results, we can compare our setup to the original reference game where there is a signal that uniquely refers to the distractor, allowing for pragmatic reasoning \citep{frank2012predicting}. We do not have such a signal in the current setup: all target features are equally overloaded. As a result, adding a layer of recursion to the pragmatic listener does not help reduce ambiguity. The only differentiating factor is expected individual utility. When the pragmatic speaker considers a signal, both literally true referents are equally likely interpretations and weighted equally regardless of the cost to reach of them. Since the optimal signaler also referred to the critical item, which was more costly to reach when the barrier was near the receiver, the expected utility of saying the optimal signal was lower.

When the task became more difficult, we did not see a significant effect of barrier location on signaling behavior. To solve the Difficult trials, both pragmatic reasoning and joint utility are required. One possible explanation for the non-significant result is that this condition is too complicated since previous human experiments on pragmatics usually contain only three referents \citep{degen2012optimal, qing2015variations, vogel2013implicatures}, but the current study has five items. More alternatives introduce more distractors, which may add difficulty to the pragmatic reasoning process. In addition, the integration of joint utility consideration and pragmatic reasoning may be more complicated than we thought. How people perceive the visual context and which items people pay first attention to may play a role in communication \citep{chun1998contextual}. Nevertheless, we still see support for joint utility reasoning making the task easier through the significant decrease of reaction time. Looking at the simulations from RSA model, as in the Simple trials, one explanation to the opposite results is that the critical item was more costly to reach. Since the expected utility of sending one signal depended on the cost of reaching all three literally true referents, it was sometimes lower than going itself or sending a signal that unambiguously referred to a wrong item but with certainty and low cost. This explains why we saw more "Do"s and "Irrational"s.

This current study provides support for joint utility reasoning in communication under a cooperative setting. Our future research aims to build a model based on joint utility calculus to capture human's signaling behavior. Moreover, we are interested in examining the effect of introducing more layers to our current RSA model and see if it will be able to solve the difficult trials. 

\bibliographystyle{apalike}
\bibliography{references}

\appendix

\section{Appendix}

\subsection{Participants}
\label{participants}
Twenty-eight University of California, Los Angeles undergraduates participated in this online study for extra credit. Participants could also claim an additional monetary reward (maximum of \$5.25) based on their performance in the task. The experiment was performed in accordance with guidelines and regulations approved by the institutional review board IRB\#19-001990.

\subsection{Stimuli}
\label{stimuli}
The experiment was completely online. Participants were able to access the experiment on their PC or laptop. Each grid in the grid world was 50 px * 50 px. The experiment can be accessed at \url{https://sy.cvls.online/signal_expt/?sonacode=123}.

\subsection{Procedure}
\label{procedure}
Participants entered the experiment by opening the link to the experiment on their own devices. Participants first read the instructions, with trials for them to try the features of going  themselves and sending a signal to the receiver. The participants could go back and forth through the instructions to make sure that they fully understood the task. After the instructions, participants completed a quiz question that tested their knowledge of the task. Participants were not allowed to proceed until they gave the correct answer. Participants would then review the consent form to proceed.

In each trial, participants could choose to go to the item themselves by clicking on the “do” button or to send a signal by clicking on a button labeled with a single feature of an item. They then watched the signaler or the receiver walk on the grids and reach an item. The receiver would wait for a random amount of time sampled based on an exponential distribution to simulate a human taking some time to think before it started to move. Once either the signaler or the receiver reached an item, a review box would pop up, showing the cost and reward of this trial, and updating the total bonus received. Participants went through 10 practice trials in which there was feedback in the review box if there was a better decision to maximize utility. In the experimental trials, there was no feedback. A debriefing survey was given after all experimental trials.

\subsection{Data cleaning}
\label{dataCleaning}
Before analyzing the data, we dropped participants and trials that failed our pre-determined criteria. Seven participants were dropped. (1 participant unfinished; 1 participant failed the instruction quiz more than twice; 3 participants self-reported that they were not serious; 4 participants failed to maximize utility in more than 25\% of the control condition, among which 2 participants self-reported being not serious; 0 participants gave the same response in more than 25\% of the total trials consecutively; 0 participants had a total utility not better than random behavior.) After dropping the participants, 56 paired trials were dropped because at least one trial in each pair had a response time that was more than three standard deviations away from the mean of the trials with the same barrier location under the same condition. Personal identifiable information for offering credits or monetary rewards was removed before analysis.

% \subsection{Results from RSA simulation}
% \begin{table}[ht]
%   \caption{Results from RSA simulation}

%   \centering
%   \resizebox{\columnwidth}{!}{\begin{tabular}{lllllllllllll}
%     \toprule
%     & \multicolumn{6}{c}{Simple} & \multicolumn{6}{c}{Difficult}\\
%     & \multicolumn{3}{c}{R} & \multicolumn{3}{c}{S} & \multicolumn{3}{c}{R} & \multicolumn{3}{c}{S}\\
%     \cmidrule{2-13}
%     & LitS & PragL & PragS & LitS & PragL & PragS & LitS & PragL & PragS & LitS & PragL & PragS\\
%     \midrule
%     Unique     & 0.5 & / & \num{3.33 e -4} & 0.5 & / & 0.176 
%     & 0.5 & / & 0.0118 & 0.5 & / & 0.277\\
%     Nonunique  & 0.5 & / & 0.999 & 0.5 & / & 0.824
%     & 0.5 & / & 0.563 & 0.5 & / & 0.304\\
%     Do         & 0   & / & \num{6.67 e -4} & 0   & / & \num{6.67 e -4}
%     & 0 & / & 0.330 & 0 & / & 0.330\\
%     Irrational & 0   & / & 0 & 0   & / & 0
%     & 0 & / & 0.0943 & 0 & / & 0.090\\
%     \\
%     Correct target   & / & 0.5 & / & / & 0.5 & /
%     & / & 0.953 & / & / & 0.953 & /\\
%     Incorrect target & / & 0.5 & / & / & 0.5 & /
%     & / & 0.047 & / & / & 0.047 & /\\
%     \bottomrule
%   \end{tabular}}
% \end{table}

\end{document}